# Efficient Super-Peer-Based Queries Routing: Simulation and Evaluation

Anis ISMAIL & Mohamed QUAFAFOU
LSIS, Domaine Universitaire de Saint-Jérôme,
Avenue Escadrille Normandie-Niemen
13397 MARSEILLE CEDEX 20
Email: {anis.ismail, mohamed.quafafou}@univmed.fr

Gilles Nachouki
LINA Laboratory
2 rue de la Houssinière
44322 Nantes cedex 03
Gilles.Nachouki@univnantes.Fr

Mohammad HAJJAR
Université Libanaise, Institut Universitaire de Technologie
Saida – LIBAN
m_hajjar@ul.edu.lb

*Abstract*—**Peer-to-peer (P2P) Data-sharing systems now generate a significant portion of internet traffic. P2P systems have emerged as a popular way to share huge volumes of data. Requirements for widely distributed information systems supporting virtual organizations have given rise to a new category of P2P systems called schema-based. In such systems each peer is a database management system in itself, ex-posing its own schema. A fundamental problem that confronts peer-to-peer applications is the efficient location of the node that stores a desired data item.
In such settings, the main objective is the efficient search across peer databases by processing each incoming query without overly consuming bandwidth. The usability of these systems depends on effective techniques to find and retrieve data; however, efficient and effective routing of content-based queries is an emerging problem in P2P networks. In this paper, we propose an architecture, based on (super-)peers, and we focus on query routing. Our approach considers that (super-)Peers having similar interests are grouped together for an efficient query routing method. In such groups, called Knowledge-Super-Peers (KSP), super-peers submit queries that are often processed by members of this group. A KSP is a specific super-peer which contains knowledge about: 1. its super-peers and 2. The others super-peers. Knowledge is extracted by using data mining techniques (e.g. decision tree algorithms) starting from queries of peers that transit on the network. The advantage of this distributed knowledge is that, it avoids to making semantic mapping, between heterogeneous data sources owned by (super-)peers, each time the system decides to route query to other (super-)peers. The set of KSP improves the robustness in queries routing mechanism and scalability in P2P Network. Compared with a baseline approach, our proposal shows a better performance with respect to important criteria such as response time, precision and recall.**

*Index Terms*—**Peer-to-peer, Query Routing, Knowledge-Super-Peers, Data Mining, Scalability.**

## I. Introduction

Peer-to-peer (P2P) systems have recently become a popular medium through which to share huge amounts of data. Because P2P systems distribute the main costs of sharing data – disk space for storing files and bandwidth for transferring them – across the peers in the network, they have been able to scale without the need for powerful, expensive servers. The key to the usability of a data-sharing P2P system, and one of the most challenging design aspects, is efficient techniques for search, route queries and retrieval of data. The major problem in such networks is query routing, i.e. deciding to which other (super-)peers the query has to be sent for high efficiency and effectiveness. The tradition P2P systems offer support for richer queries than just search by identifier, such as keyword search with regular expressions. Search techniques for these systems must therefore operate under a different set of constraints than techniques developed for persistent storage utilities.

However, such systems that broadcast all queries to all peers suffer from limited efficiency and scalability, and are difficult to locate files which result also in much network traffic and low recall/precision. In hybrid P2P systems [1][2], composed of (super-)peers, when a peer submits a query, this peer becomes the source of this query. Then the query is transmitted to its super-peer (SP). The routing policy in use determines the relevant neighbors (i.e. SP) quickly, based on semantic mappings between schemas of (super-)peers, and to which neighbors, the query is sent. When a SP receives a query, it will process the query over its local collection of data sources of different peers and sends the query to relevant neighbors (SP) for processing [24]. If any results are found, the SP will send a single response message back to the query source. Another important aspect of the user experience is how long the user must wait for the results to arrive. This is due to a large part of the mediation process which remains difficult to realize in such a context when the number of (super-)peers increases. Response times tend to be slow in hybrid P2P networks, since the query travel through several SP in the network and whenever the SP is forced to look for connections (i.e. mappings) in order to route the query.

Satisfaction time is simply the time that has elapsed from when the query is first submitted by the user, to the other users that contain the most relevant answers in a fast and efficient way, until the user receives the overall results. This also is the main challenge of information retrieval in Peer-to-Peer networks [12].

In this paper, we present an approach for efficient queries routing. The important advantage of this approach is scalability. Our system is designed to efficiently support content-based searching. Our main goal is to reduce the processing of queries at the SP level to predict others relevant SP to receive and process such queries. Our proposed method focus on how the query is routed to relevant Peers with minimum query processing in order to improve answering time of the queries.

Our approach consists of grouping together (super-)Peers that have similar themes for an efficient query routing method. Each obtained group, called Knowledge Super-Peers (KSP), contains communities, composed of super-peers (the responsible of communities) and their corresponding peers (the members), that submitted queries that are often processed by members of this group (after grouping). Each KSP operates with an index that, obtained by applying decision tree algorithms, keep track of where contents concerning a query are located : when a KSP receives a query from a Super-Peer (in his group), it consults directly its index (without making any mappings) in order to determine: 1. in his group all super-peers (or communities)that are able to answer this query and 2. in other groups (i.e. other KSP) all super-peers which are relevant to this query. In this paper, we do not care how we get the different groups of SP but we focus only on the Super-Peer based routing protocol of users's queries.

The following section recalls briefly principal concepts of P2P networks and shows the context of our work. Section 3 presents the baseline algorithm of queries routing in hybrid P2P systems. Section 4, introduces the Knowledge super-peer (KSP) network. Section 5 presents the semantic routing of queries algorithm. Section 6 presents Experiments and Evaluations. Section 7 shows the results of our experiments. In Section 8, we present the conclusion and future works.

II. BACKGROUND

A. Basic notions

A Peer is an autonomous entity with a capacity of storage and data processing. In a computer network, a Peer may act as a client or as a server. A P2P is a set of autonomous and self-organized peers (P), connected together through a computer network. The purpose of a P2P network is the sharing of resources (files, databases) distributed on peers by avoiding the appearance of a peer as a central server in this network. We note: P2P = (P, U), P is the set of peers and U represents links (overlay connections) between two peers Pi and Pj, $U \subseteq P \times P$.

The hybrid P2P (P2Ph) (See Figure 1) network that we consider in this paper includes sets of peers (P) and super-peers (SP). We note : P2Ph = (P $\cup$ SP, K), where P is the set of peers, SP is the set of super-peers and K is the set of overlay links expressed under the format of pairs : (Pi, SPj) or (SPj ,SPk) which respectively link a Peer Pi to a Super-Peer SPj or a Super-Peer SPj to one or several super-peers SPk.

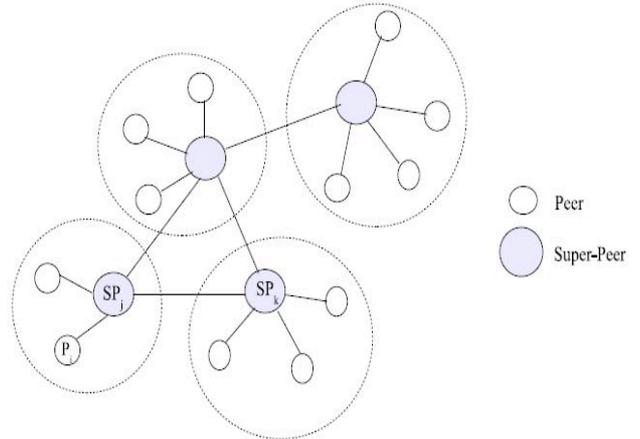

Figure 1. Hybrid network (P2Ph).

A PDMS (Peer Data Management System) combines P2P systems and databases systems. The PDMS that we are considering is a scale hybrid system P2Ph. Each peer is supposed to hold a database (or an XML document, etc.) with a data schema. Each Super-Peer provides a theme (a semantic domain, a subject, or an idea) representing special interest to a group of peers. The themes are not necessarily separated; they are described by super-peers, with the three following manufacturers:

– A concept is a collection of individuals that constitute the entities of the modeled domain. The concepts can be compared to the notion of class (i.e. object model) or type of entity in the conceptual models (i.e. Entity/Relationship).

– A role is a binary relationship between concepts. Roles are used to specify properties of instances and are compared to the notion of attributes in the conceptual models. A role is viewed as a function linking a concept (called domain) to another concept (known as co-domain).

– Specialization (IsA) starts from a specific concept to a more general concept. It is transitive and asymmetric and defines a hierarchy between concepts it connects.

We note R the set of relations reduced in this paper to two relations that are {Role; IsA} and PDMS={PS $\cup$ SPT, D , K} where PS represents all the peers of the network with their data schemas S={S1, …., Sp}. A peer is connected to the network with only one data schema. K is the set of overlay links between (super-)peers. Each peer P $\in$ PS is doted of a Data Management System (denoted DMS) able to manage their data.

T={T1,…., Tk} represents the interest themes published by super-peers SP through the network. In our case, each super-peer publishes only one theme and peers expresses that are interested by one or several theme(s) in T. The themes are not disjoints: two super-peers can publish the same concepts or roles with distinct structures and/or don't use the same vocabulary. D = {D1, …., Dk} describes the themes in the set of T: Dj describes the theme Tj specifying the set of concepts and their relationships (see figure 2).

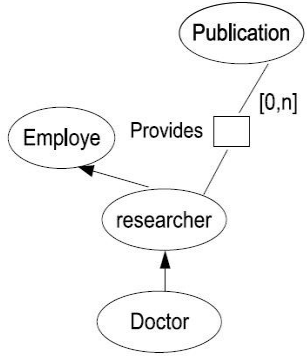

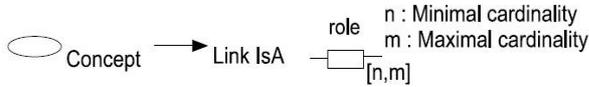

Figure 2. A part of Hospital theme published by SPj

### B. Expertise, Mapping and Communities

At this step, we consider only data models supported by peers. We distinguish the three following data models, the best known: relational, XML and object. An expertise is defined, in our case, as (a part of) the data schema, expressed with one of the three data models cited above, possessed and published by a Peer in order to share its data with other peers. To facilitate the reconciliation, between the data schema of the Peer and the theme described by a Super-Peer, two measures were taken: 1. the expertise of a Peer is expressed with the language of its Super-Peer (i.e. concept, role and IsA); 2. The expertise of a Peer is expressed under the format of couple of elements, satisfying the following condition:

EXP (Pi) = { $\theta$(si; sj) $\in$ SP | (si; sj) $\wedge$ $\theta$ $\in$ R}

Example: In this example, we express the schema of the Peer Pi with the language of description of the Super-Peer. Among the concepts of this schema of the Peer we found: Employee, Publication, Researcher and Doctor. Some links are established between the concepts: Employee and Researcher (see Figure 2). This link expresses that a Researcher is an Employee. The expertise of Pi is given as follows:

EXP (Pi) = {IsA(Researcher;Employee);provides (Researcher; Publication); IsA(Doctor;Researcher)}.

In our context, mapping is an important process in order to share data between peers. Two levels of mapping are distinguished: the first level is to share data between peers, it is important to search for connections between expertise of peers and the description of themes provided by super-peers. The second level is to process users queries, its important to search for connections between the subject of a query (detailed below) and the expertise of each (super-)peers in order to know its capacity to response to this query. Let S1, the expertise of a peer and S2 the theme proposed by the Super-Peer of its community. The search for correspondence between S1 and S2 is to find for each concept or role in S1 (or S2) a correspondent in S2 (or S1) which is the nearest semantically. We can define the concept of mapping (Map) between schemas as follows:

Map: S1 $\rightarrow$ S2   Map(es1) = es2    if     (1)

Sim(es1; es2) > acceptable-threshold

Where es1: entity of schema S1; es2: entity of schema S2;

Sim(es1; es2) is a function, that measure the similarity between two entities es1 et es2, given as follows:

Sim: S1xS2 $\rightarrow$ [0; 1]        (2)

We distinguish two particular cases: Sim(es1; es2) = 1 describes two similar entities; Sim(es1; es2) = 0 describes two distinct entities.

We introduce the two concepts, Semantic Intra-Community and Semantic Inter-Community. A Semantic Intra-Community is an interest community in which mappings between peers, members of this community, and the Super-Peer, responsible of this community, are established. A Semantic Inter-Community is a set of semantic Intra-community in which mappings between Super-peers of these communities are established.

We note Semantic Intra-Community ($SI_a^jC$) and Semantic Inter-Community ($SI_a^jC$) number j as follows:

$SI_a^jC$ = (PS $\cup$ SPTj,Dj , EXP(PS), Kj ;RSCj)    (3)

$SI_a^jC$ = ($SI_a^jC$, RSIj,1, …, RSIj,k), k $\neq$ j    (4)

where PS $\subseteq$ P is a subset of peers having the same center of interest Tj , EXP (PS) is the set of expertise of peers interested by this theme and joined to this community, SP$_{Tj,Dj}$ (belong to SP) is the Super-Peer responsible of the community j which are joined by peers (i.e. a Peer of a community may request to join several communities if the user thinks that his theme of interest is in the intersection of several communities), Dj represents the description of the theme Tj provided by the Super-Peer. Kj $\subseteq$ K is the set of overlay links between the super-peer SPTj,Dj and the peers connected to it union the set of overlay links between SPTj,Dj and Super-Peers SPTk,Dk, k $\neq$ j, RSCj is the semantic Intra-Community between the super-peer SP$_{Tj,Dj}$ and the peers inside this community. RSIj,k is the semantic Inter-community concerning the links found between the description of the theme Dj of the Super-peer SPTj,Dj , with the description Dk of each super-peer SPTk,Dk, k ≠ j). Finally, we introduce a Semantic Overlay Network (SON) represented by the union of all the semantic networks of intra-communities and inter-communities. A SON is noted as follows:

$$SON = \bigcup_{j=1}^{|T|} (SI_e^j C) \quad (5)$$

Where T represents the total number of super-peers in the network. Next section presents the query routing algorithm (our baseline approach).

## III. SEMANTIC QUERIES ROUTING - BASELINE

### A. Network Configuration

A new Peer Pj advertises its expertise by sending, to its Super-Peer, a domain advertisement DAj = (PID; $E_{XP}^j$, Tj ; Ɛacc; TTL) containing the Peer ID denoted PID, the suggested expertise $E_{XP}^j$, the topic area of interest Tj , the minimum semantic similarity value (Ɛacc) required to establish semantic mapping between the suggested expertise $E_{XP}^j$ and the theme of its Super-Peer. When receiving an expertise $E_{XP}^j$, a Super-Peer SPa invokes the semantic matching process to find mappings between its suggested schema and the received expertise.

### B. Baseline approach

A Peer submits its query on its local data schema. This query is sent to its Super-Peer responsible for the community (see Figure 3).

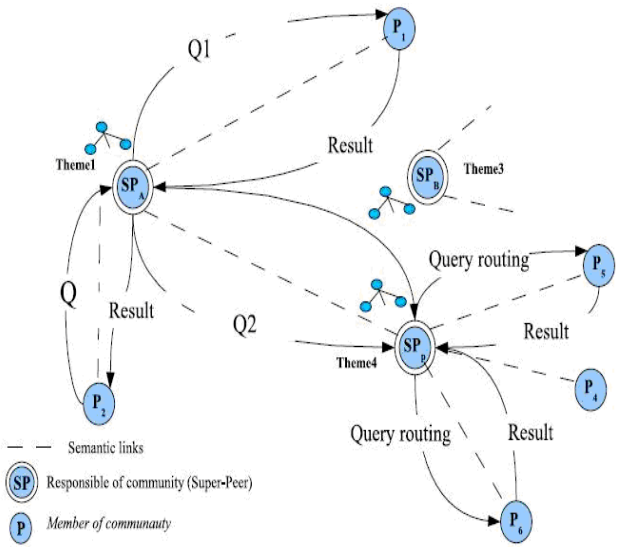

Figure 3. Network configuration and query routing (baseline).

The Super-Peer in its turn suggests, based on the index obtained by the process of mediation (first level), the peers of his community or the other super-peers that are able to treat this query. Each submitted query received by a Super-Peer, is processed by searching connections (second level of mappings) between the subject of this query and expertise of peers (of the same community) or the description of themes of other Super-peers. In its turn, a super-peer from the nearby community, having received this request, researches among peers (in his community) that are able to answer this query. The major problem of this approach is the mediation at the two levels cited above: if we take thousands of peers or super-peers this approach can not be scaled due to the mappings at both levels. The followings sections describe our approach in order to avoid super-peer, when it's too busy to treat all users' queries, to process the second level of mapping. This approach improves response times of queries and scalability in P2Ph context by restructuring the network dynamically. To do that, we introduce the concept of Knowledge-Super-Peer (KSP).

## IV. KNOWLEDGE -SUPER-PEER

A Knowledge-Super-Peer (KSP) network is a semantic sub-network of Overlay Network (SON). The KSP number j is defined as follows:

$$KSPj = \bigcup_{l=1}^{|M|} (SI_e^l C) \quad |M| \leq |T| \quad (4)$$

Where M is the number of Super-Peer in KSPj and |M| ≤ |T| (total number of super-peers). $SI_e^l C$ is the Semantic Inter-Community of the super-peer number l. Two fundamental properties are derived from KSP:

$$KSPi \cup KSPj = SON, i \neq j \quad (6)$$

A Knowledge-Super-Peer is represented physically with a specific Peer. This Peer, representing the Knowledge-Super-Peer number j, is noted as follows:

KSPj = (PS $\cup$ SPT J, DJ, EXP (PS), Kj, RSCJ, RSIJ, INDj) where PS $\subseteq$ P is a subset of peers having very close center of interests denoted T J = {T1,…, Ts}, EXP (PS) is the set of expertise of peers interested by at least one of themes in T J, SP$_{T J, DJ}$ (belong to SP) is the set of super-peers responsible of communities which have very close domain interests, DJ = {D1, …, Ds} represents the description of themes in T J (DJ describes TJ). Kj $\subseteq$ K is the set of overlay links between each super-peer SPTj, Dj $\in$ SPT J, DJ and 1. The peers connected to it (within its community); 2. The other super-peers; 3. The Knowledge-Super-Peer KSPj itself. RSCJ is the set of semantic Intra-Community of the super-peers $\in$ SPTJ, DJ . RSIJ is the set of semantic Inter-Community for each super-peer in SPTJ, DJ. INDj is the index obtained using a decision tree algorithm to identify directly the most relevant (super-)peers, without going through mappings, to provide good results when a query is submitted by a peer.

Our proposed System (See Figure 4) is an hybrid P2P system based on an organization of peers around super-peers according to their proposed themes, where super-peers are connected to a Knowledge-super-peer (KSP), the engine that specifies the super-peers having peers which may have relevant data to answer queries with minimum query tasks and, by consequence, improve answering time of the queries. The super-peer architecture allows the heterogeneity of peers by assigning more responsibility to peers able to assume them.

Therefore, certain peers, called Knowledge super-peers, have an additional computing power and greater bandwidth, resources and performing administrative tasks. They are responsible of routing queries to relevant super-peers, allowing not only to reduce efforts of compilation of queries but also to prevent the spread of queries in the network. In each community, there is a super-peer connected to a Knowledge super-peer where we have an index to identify super-peers that are most relevant to provide good results of queries.

The building block (KSP) of the current P2P systems in the architecture (Distributed Knowledge - DK) is the notion of a super-peer-group, or a number of nodes (super-peer) that participate with each other for a common purpose to minimize the load in the KSP. Example : In this example we explain the query routing using KSP (Figure 4), A Peer P2 sends a query Q2 to his SP (SPA) that in its turn sends this query to KSP that belong to and also to peers of his community that are able to answer this query. This KSP analyzes the query to find the other SP using decision tree to send this query. Finally, the results will be sent to P2.

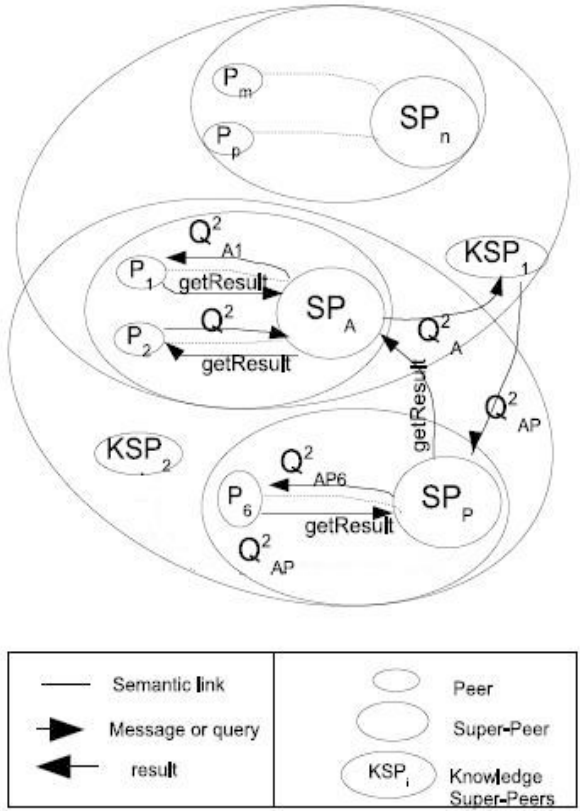

Figure 4. Network configuration and query routing (KSP approach).

V. SEMANTIC QUERY ROUTING ALGORITHM

Our algorithm of semantic query routing is composed of two stages: the semantic routing algorithm (Algorithm 1) of the baseline approach exploits the expertise of (super-)Peers and the two levels of mappings in order to forward a query q to only relevant Super-Peers. Each Super-Peer in its turn forwards this query to relevant Peers in its community. The followings sub-steps are necessary in order to process the query: 1. Extract the subject of this query; 2. select, by the super-peer, the most relevant peers for the query and the other super-peers (by matching the subject of the query to the set of expertise of peers or to the themes of super-peers). The selection is based on a function that measures the capacity of a peer or a super-peer on answering a given query; 3. Once the set of relevant (super-)peers has been identified, the super-peer sends the query to those promising peers or super-peers closed to them by using their ID, IP addresses and the underlying physical network. The advantage of this step is that it permits us, for the second step, to collect information about the queries received by super-peers and the relevant super(-peers) selected in order to process it. The second algorithm exploits the Knowledge-super-peers (KSP) network.

| Algorithm1: Baseline algorithm: BL(Q,SP) |
|---|
| Inprut: Q: Query |
|         SP: Super-Peer of P |
| Output: $SR_Q$ : Set of answers of Q |
| Variables: PSet: Set of Peers |
| NP: Neighbors of SP (Set of Super-peers) |
| $SR_Q = \phi$ |
| 1:   Pset = $Capacity_{CM^{SP/SP}}(Q) > \in_{acc}$ |
| **repeat** |
| 3:    $SP_Q$ = get(s $\in$ PSet); |
| 4:    Remove $SP_Q$ from Pset; |
| 5:    $SR_Q = SR_Q \cup$ Query($SP_Q$); |
| 6: |
| **Until** (PSet= $\phi$ ) |
| **repeat** |
| 10:    $SP_Q = Capacity_{CM^{SP/SP}}(Q) > \in_{acc}$ |
| 11:    Remove $SP_Q$ from NP; |
| 12:    $SR_Q = SR_Q \cup$ BL(Q,$SP_Q$); |
| **Until**(PSet= $\phi$ ) |
| Return($SR_Q$ ); |

| Algorithm2: Knowledge based algorithm KB(Q,SP) |
|---|
| Input: Q: Query |
|         SP: Super-Peer of P |
| Output: $SR_Q$ : Set of answers of Q |
| Variables: $T_{SP}$: decision Tree of SP |
| NP: Neighbors of SP (Set of Super-peers) |
| $SR_Q = \phi$ |
| Pset = Select(p $\in$ SP) |
| **repeat** |
|     $SP_Q$ = get(s $\in$ PSet); |
|     Remove $SP_Q$ from Pset; |
|     $SR_Q = SR_Q \cup$ Query($SP_Q$); |
| **Until** (PSet= $\phi$ ) |
| $SP_Q = T_{SP}(Q)$; |
| $SR_Q = SR_Q \cup$ Query($SP_Q$); |
| Return($SR_Q$ ); |

This algorithm (algorithm 2) is very useful when the performance of the system is low.

This step runs in three stages: 1. the super-peer sends the query directly to its Knowledge super-peer; 2. the Knowledge super-peer identifies (without make mapping) the relevant KSP for this query and their super-peers by consulting its index IND (obtained by applying decision tree algorithms); 3. Each selected super-peer sends the query to relevant Peers; 4. The final result of selected peers is returned.

## VI. SIMULATOR ARCHITECTURE

The simulation is a technique for modeling the real world [27]. It can represent the operation of a system consisting of various activity centers, to reveal the characteristics of them and the interactions between them, to describe the movement of the various subjects treated by these processes and finally to observe the behavior of the system as a whole and its evolution over time. The discrete event simulation can help understand the behavior of the system. Several research projects such as Freenet [28] and Anthill [29] have used simulation in order to show their performance. The discrete event simulation allows observing the behavior of the system. The model has a state described by variables that define completely the characteristics of the system [30].

There are several peer-to-peer simulators available: P2PSim [31] is a discrete event simulator for structured overlay networks written in C++. It comes with seven peer-to-peer protocols implemented including the more recent protocols Koorde [34] and Kademlia [35]. There are a number of different underlying network models, all of them, however, on a rather abstract level of detail, making it hard to simulate the dedicated overlay devices in the access networks mentioned above. P2PSim is largely undocumented and therefore hard to extend.

OverlayWeaver [36] is a peer-to-peer overlay construction toolkit written in Java which can be used for easy development and testing of new overlay protocols and applications. The toolkit contains a so-called Distributed Environment Emulator which invokes and hosts multiple instances of Java applications on a single computer. This allows the simulation of up to 4,000 nodes. Since simulations have to be run in real-time and there is no statistical output, its use as an overlay network simulator is very limited.

PlanetSim [37] is an object-oriented simulation framework for overlay networks and services written in Java. It has a well-structured and modular architecture and makes use of the Common API [38]. In addition to the overlay protocols Chord [39] and Symphony [40] there are several services like CAST and DHT available on application layer. PlanetSim offers only limited support to collect statistics and has a very simplified underlying network layer without consideration of bandwidth and latency costs. This makes it difficult to simulate heterogeneous access networks and terminal mobility. It is possible to visualize the overlay topology at the end of a simulation run, but there is no interactive GUI.

A more comprehensive survey of peer-to-peer network simulators can be found in [41], where the authors show that most available peer-to-peer network simulators have several major drawbacks limiting them in use for research projects.

The state model is often encapsulated in a set of entities (objects in object-oriented programming). The discrete event changing the system state that occurs at different points in time (as opposed to the continuous change of states). Events may trigger new events. Statistical variables then define performance measures relevant to the user.

In this section, we present our P2P network simulator domain-based semantics. The proposed simulator is then extended to search for information in a P2P context. The simulation process that we present in Figure 5 consists of five main stages, each supporting a set of generic functions:

1. **Initialization**: The initialization phase permits to acquire the user preferences. These preferences mainly concern the number of peers, super peers, the various fields (super-) peers and the choice of strategy (semantic-oriented or knowledge-based ties) to be used during the "Management queries. Based on user preferences, a set of parameters common to all other strategies is generated. These parameters are mostly the identification of areas of expertise and the generation of super-peers.

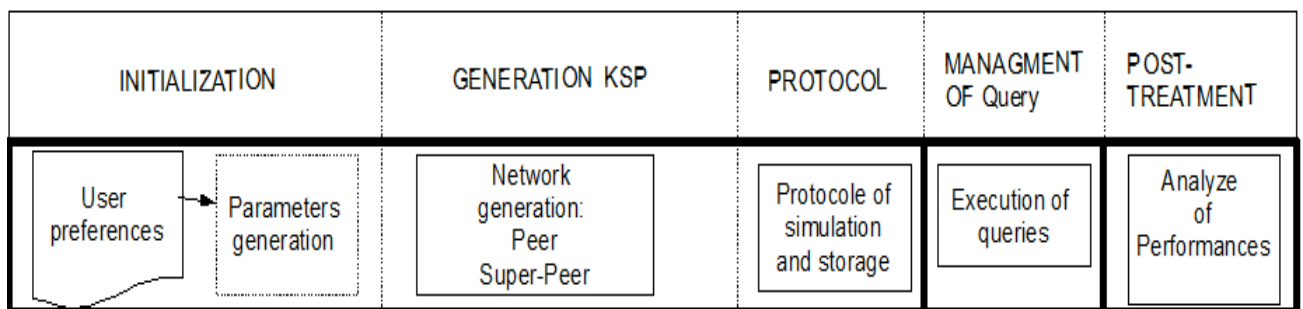

Figure 5. Simulation process

2. **Generation KSP**: The generation phase KSP network is an important step in the process of simulating semantics P2P network. This phase permits to construct and simulate KSP networks accordance architectures presented in this paper.

3. **Protocol**: the protocol allows specifying some basic rules necessary for the proper functioning of the simulator. On Stage "Managing Queries", the simulator must know which method to adopt to send the queries: for example, a first method is to generate a query in pairs; queries are generated and sent along with the super- peers for processing. Another method is to generate multiple requests (a number of query i randomly chosen between 1 and N) per pair. Regarding the management of the network, the simulator needs to have information on domain-groups: they can be dynamic, for example, on one hand a super-peer in a domain-group may transfer to another domain-group if it does anything (knowledge) to this domain-group and on other hand, a super-peer may leave the network completely so. Regarding the management of knowledge in a domain-group, we can distinguish cases where knowledge at the domain-group level can be static or dynamic. Knowledge dynamics are updated periodically by the relevant domain-group.

4. **Management of Query**: This phase involves generating a plan for routing queries. The generated plan is built by one of the strategies described in this paper: semantics or domain-groups. The KSP approach is a hybrid approach since it is based on two strategies: knowledge-oriented or minimal ties (to search for relevant super-peers can answer a query) and semantic (search within each Super-Pair relevant peer that can respond to this query).

5. **Post-treatment**: This phase involves defining the types of expected results and analyzes the performance of each simulation performed.

*A. Semantic Aproach*

In the semantic approach (Figure 6) several parameters are needed to build the semantic SON. Among these parameters, we include the number of peers and super-peers that make up our network areas of (Super-) peers; different thresholds: 1. A level of correspondence (mappings) deemed acceptable by the (Super-) peers; 2. An acceptable threshold for establishing trust between two super-peers and 3. The ability of a (Super-)peer to process a request. These parameters are common for different strategies.

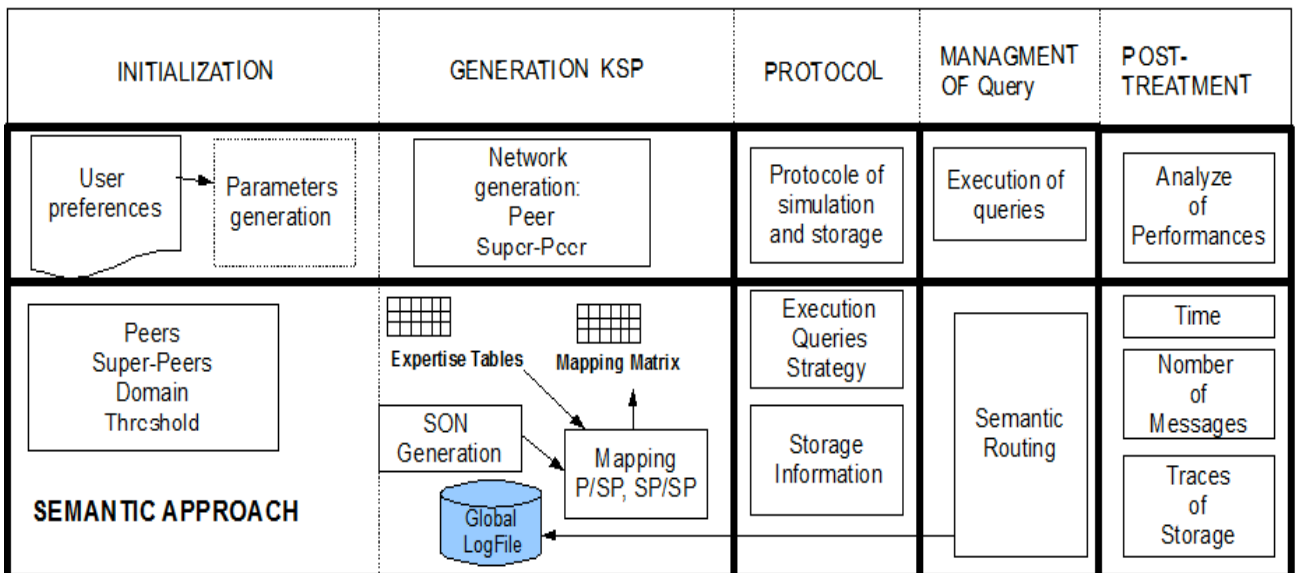

Figure 6. Simulation process – Semantic Approach

The algorithm 3 initializes the system with the generation of areas and expertise of the super-peers:

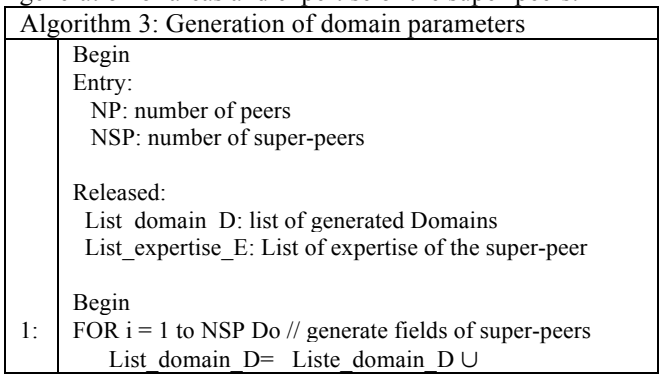

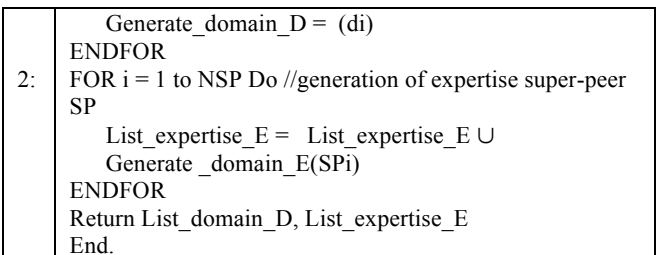

| Algorithm 3: Generation of domain parameters |
|---|
| Begin<br>Entry:<br>  NP: number of peers<br>  NSP: number of super-peers<br><br>Released:<br>  List_domain_D: list of generated Domains<br>  List_expertise_E: List of expertise of the super-peer<br><br>Begin<br>1:  FOR i = 1 to NSP Do // generate fields of super-peers<br>    List_domain_D= Liste_domain_D ∪<br>    Generate_domain_D = (di)<br>  ENDFOR<br>2:  FOR i = 1 to NSP Do //generation of expertise super-peer SP<br>    List_expertise_E = List_expertise_E ∪<br>    Generate _domain_E(SPi)<br>  ENDFOR<br>  Return List_domain_D, List_expertise_E<br>End. |

The size of the network being defined by the number of peers NP and super-peers NSP that are given by the user. For each super-peer I, we generate using the function generate_domain a label di which is the name of the domain represented by the super-peer i (step 1 of algorithm 3). This generation respects the following condition: two super-peers can not be assigned to the same domain. Then it generates, in step 2, the expertise of each super-peer represented as a set of couple (x, y). We note χ (X) the expertise of the super-peer X.

To generate the SON networks, we start building the correspondences (mapping) between the super-peers. Then, we generate peers, their expertise and we implement the connections between peers/super-peers. It begins by calculating the correspondence (mapping) between the semantic super-peers (algorithm 4). For this, we represent the expertise of super-peer by an expertise table (ExpTabSP) of super-peers. To simulate this calculation, a super-peer selects randomly a number of super-peers of the network to consider them as friends, and then duplicate some elements of its expertise in the expertise of his friends. The number of duplicate elements has to be selected in order to ensure the existence of mapping between a super-peer and his friends.

| Algorithm 4: Generation of SON Network (SP/SP) | |
|---|---|
| | Pre-condition: SPi is a new super-peer in the network |
| | Begin |
| | Entry: |
| |   ExpTabSP table d'expertise SP |
| |   CorMatSPSP: Matrix correlation SP/SP |
| | Output: |
| |   TFI: table of friends of the super-peer SPi (initially empty) |
| |   CorMatSPSPi: Changing the correlation matrix |
| | TFI = Select_ friend (SPi) // selects SPi Friends |
| | FOR each  super-peer SPj ∈ TFi  Do |
| 1: | T = select_expertise (ExpTabSPi) // selection elements exp. SPi |
| | Send (SPi, T, SPj) / / Send selected elements to SPj |
| 2: | Addition (SPj ExpTabSPj, T) // SPj ExpTabSPj addition to the elements of T |
| | Addition (SPj, TFj, T, SPi) // SPj has a new friend SPi |
| 3: | Update (SPi, CorMatSP) // update mapping SPi |
| 4: | Update (SPj, CorMatSP) // update mapping SPj |
| | ENDFOR |
| | END |

| Algorithm 5: Generation of SON Network (P/SP) | |
|---|---|
| | Begin |
| | Entry: |
| |   NP: number of peers |
| |   MIN: minimum size of the expertise of a peer |
| | Output: |
| |   ExpTab table of expertise of P |
| |   CorMatSPP: correspondence Matrix of super-peer/peer |
| 1: | FOR i = 1 to NP do // generate expertise of peers P |
| |   List_expertise=generate_expertise (Pi, SPi, MIN) |
| |   ExpTab = store (List_expertise) |
| |   Storage(ExpTab, SPi) |
| | ENDFOR |
| 2: | Create nodes (SP) // creation of SON Network |
| | Create nodes (P) |
| 3: | CorMatSPP= Create_Correspondance = (P, SP) //create link peer/super-peer |
| | End. |

At this level, the SON network is built; it remains to clarify the evolution of its architecture based on the dynamics of peers and super-peers.

Trust between two super-peers depends on the number of semantic links connecting them. The trust is useful where a super-peer SP leaves the network: peers attached to SP will then be attached to the super-peer with the highest degree of trust with SP.

We consider that the queries are expressed in the simulator in the form of elements of expertise: $\bigwedge_{i=1}^{n} (p_i.q_i)$ where $p_i.q_i$ are easily comparable with the components of expertise of peers and super peers. It is considered that the rewritten query the user as the elements of expertise is not part of the simulator, but it is a task delegated to the mediator.

The generation of applications is ensured by peers. In fact, each peer P can generate a query by selecting elements of expertise that become components of the query Q. We say that a peer P is relevant to the query Q if the expertise of P contains at least a fraction of the components of Q. This is determined using the ability of a peer P to resolve a query Q.

So each peer generates a number N of queries that are derived from its expertise. After this phase generation of query, peers send their queries to their super-peers. Algorithm 6 shows in detail the stage for routing queries in the context of the semantic approach. In fact, step 1 show that all peers send their queries to their super-peer at time t. The super-peer that receives the query performs a local search (step 2) by considering only one pair that belongs to the domain it represents. Then, the super-peer sends the query to his friends, that can respond to the query for global search (step 4).

| Algorithm 6: Query routing, Generation of global LogFil | |
|---|---|
| | *Pre-condition*: The Queries are in the parameters file. |
| | identifier strategy (ids = 1) |
| | Begin |
| | *Input:* |
| |  *ExpTabSP:* Table of expertise associated with the super-peer SP |
| |  *ExpTabP:* Table of expertise associated with the peer P |
| |  Threshold: threshold acceptable |
| 1 : | At time t: ∀ P of the network has SP as  super-peer Do |
| |   send(P, Q, SP) // P sends  its Query Q to SP |
| 2 : | Perform local search |
| 3 : |  List_P = search(SP, *ExpTabP*, Q) //search pertinent peers |
| | While Pk ∈ List_P Do |
| |   Send(SP, Q, Pk)  // Send Q to Pk |
| 4 : | EndWhile |
| | Perform global search |
| 5 : | A = Friends(SP) / / A all the super-peer friends of  SP |
| 6 : |  While SPk ∈ A Do |
| 7 : |  List_SP  =  search(SP, *ExpTabSP*, Q)   // Search pertinents SP for Q |
| 8 : |     While SPk ∈ List_SP Do  // for all SP that can |

```
                              process the Queries
            Send(SP, Q, SP_k)  // Send Q to SP_k
                               // SP_K performs a local search
            List_P = Search(SP_k, ExpTabP, Q)
            //Search pertinents P
              While P_i ∈ List_P Do
                 Send(SP_k, Q, P_j) // Send Q to P_k
              Endwhile
           Endwhile
        Endwhile
     End
```

All queries exchanged within the network are stored in a file global LogFile. Thus, for a query Q, the file LogFile contains the following information: the identifier of the peer (P), which submitted the application, its super-peer (SP), the query (Q) itself and the super-peer which responded favorably to this request.

*B. Aproach KSP*

In this section we present an SON-KSP network based on knowledge. We begin first by simulating domain-groups oriented knowledge (Figure 6). At this level, domain-groups are built, above the previously established semantic layer, based on reliance by a member from one domain-group to another domain-group member. Indeed, a super-peer (referred to as a domain member) is free to join a domain-group if at least one member of this domain-group has given him confidence.

In this knowledge-oriented approach and to initialize the system, the user must give the acceptable threshold for establishing trust between the super-peers and must decide the dynamics of knowledge within domain-groups (refresh knowledge) then begins the generation of SON-KSP network. At this level it is to extend the SON network to SON-KSP. A domain-group is characterized by the knowledge that bears on its super-peer as well as super-peers in neighboring domain-groups. According to user preferences at this level we build one or more domain-groups. Building a domain-group center is directly from the previously built global LogFile (strategy based on the semantics). The construction of several domain-groups is mainly based on the notion of trust referred to in the preceding section. Indeed, the SON network is built from a layer dedicated to peers and another juxtaposed containing super-peers, and above a third layer was built domain-groups (SON-KSP) where each node is a domain-group.

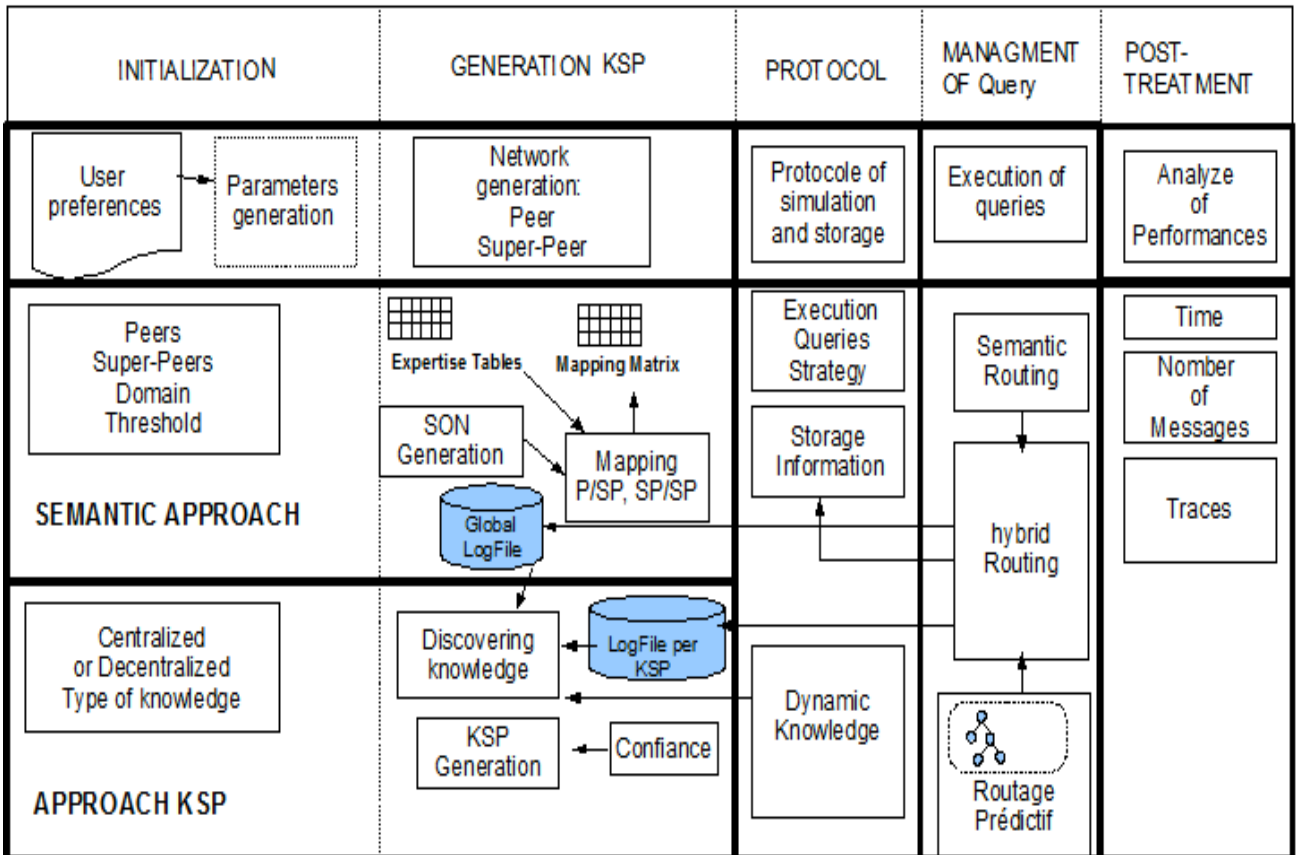

Figure 7. Simulation process – Approach KSP

The knowledge-oriented approach combines knowledge of each domain-group it owns (SON-KSP). Before extracting the knowledge of domain-group, we need to involve each domain-group its log file containing all the queries processed by one of its members (super-peer). The data contained in this file will be analyzed by domain-group using a tool data mining to extract knowledge. The role of knowledge in this context would be to predict the super-peers that may treat a given query. We express this knowledge in the form of a decision tree.

In practice, we build knowledge of different obtained domain-groups. We used the J48 algorithm implemented by WEKA and its inference methods to find the super-peers probable to treat a given query. The developed method analyzes the probability distribution in space of super-peers and keeps only those that have a nonzero probability. The traces of the simulation are stored in different files to support post-processing methods for analyzing and comparing the results of several simulations.

## VI. EXPERIMENTS AND EVALUATIONS

Decision trees represent a supervised approach of classification. We have used Weka [25] for our experiments. The most important part of the entire data mining process is preparing the input for data mining investigation.

Our P2P database contains data from more than 300 peers with 10 super-peers (data contains the keywords (composant W1...W4) search of queries (part of expertise of peers (k.f, p.i, f.p, g.h, ...) and their answers (relevant peers with their super-peer)), after a simulation in the Architecture-baseline and data Extraction and filtering to obtain the ARFF format that is input data to be injected in Weka to obtain the decision tree. Decision trees are often used in classification and prediction. It is a simple and powerful way of knowledge representation. The models produced by decision trees are represented in the form of tree structures. A component of query indicates the class of the examples. The instances are classified by sorting them down the tree from the first component of the query to other component of the query.

Decision trees represent a supervised approach of classification. Weka uses the J48 algorithm, which is Weka's implementation of C4.5 Decision tree algorithm. J48 is actually a slight improved to and the latest version of C4.5. It was the last public version of this family of algorithms before the commercial implementation C5.0 had been released. C4.5 was chosen for several reasons: it is a well-known classification algorithm; it has already been used in similar studies and it can originate easily understandable rules. J48 is the decision tree classification algorithm. It builds a decision tree model by analyzing training data, and uses this model to classify user data. Figure 8 shows the results of running J48 Decision tree algorithm. Each line represents a node in the tree (See Figure 5). The second two lines, those that start with a '|', are child nodes of the first line. In the general case, a node with one or more 'j' characters before the rule is a child node of the node that the right-most line of '|' characters terminates at, if you follow it up the page. The next part of the line declares the rule. If the expression is true for a given instance, you either classify it if the rule is followed by a semicolon and a class designation that designation becomes the classification of the rule-or, if it isn't followed by a semicolon, you continue to the next node in the tree (i.e. the first child node of the node you just evaluated the instance on). If the expression is instead false, you continue to the "sister" node of the node you just evaluated; that is, the node that has the same number of '|' characters before it and the same parent node. Nodes that generate a classification, such as composanteW1 = j.m:SP1 (50.0) are followed by a number (sometimes two) in parentheses. The first number represents how many instances in the training set are correctly classified by this node, in this case 50 are.

The second number, if it exists (if not, it is taken to be 0.0), represents the number of instances incorrectly

```
composanteW1 = k.f
|   composanteW2 = p.i: SP0 (26.0/11.0)
|   composanteW2 = f.p: SP0 (12.0)
|   composanteW2 = g.f: SP3 (26.0)
|   composanteW2 = g.h: SP0 (15.0)
composanteW1 = p.i: SP0 (50.0)
composanteW1 = f.l: SP1 (78.0/12.0)
composanteW1 = f.p: SP1 (159.0/14.0)
composanteW1 = d.o
|   composanteW4 = r.m: SP3 (38.0/16.0)
|   composanteW4 = i.c: SP3 (25.0)
|   composanteW4 = s.d: SP6 (28.0)
composanteW1 = r.m: SP5 (393.0/138.0)
composanteW1 = g.f: SP3 (46.0)
composanteW1 = i.c: SP3 (157.0/37.0)
........
```

Figure 8. Decision tree for KSP03 (for example).

classified by the node. The Classification of large data-sets is an important data mining methodology. For our purposes the most important figures here are the numbers of correctly and incorrectly classified instances. The output from the Weka program is shown in the Figure 6. In this output, the decision tree is able to classify approximately ninety two percent of the data correctly.

We describe the performance evaluation of our routing algorithm with a SimJava-based simulator [6]. All experiments were run on a machine Core 2 Duo 1.83GHZ with 4 GB RAM, 250 GB Hard disk and Windows Vista operating system. Evaluating the performance of P2P network is an important part to understand how useful it can be in the real world. As with all P2P applications, the first question is whether P2P is scalable. Our systems were evaluated with different set of parameters i.e. number of Peers, number Super-peer etc. Evaluation results were quite encouraging.

There are many dimensions in which scalability can be evaluated: one important metric is the time it takes the Answer of a given query. We run simulations on P2P network in three different sizes. Each peer sends a query to its SP that sends the query to a KSP in order to precise which Super-peer(s) can answer the given query, this in the architecture-DK.

- First one, we modified the number of Peers (300, 600,..., 5000 Peers) and Super-peers (10,12 ,14, 16, 20,..., 54) in the both Architectures to measure the execution time.

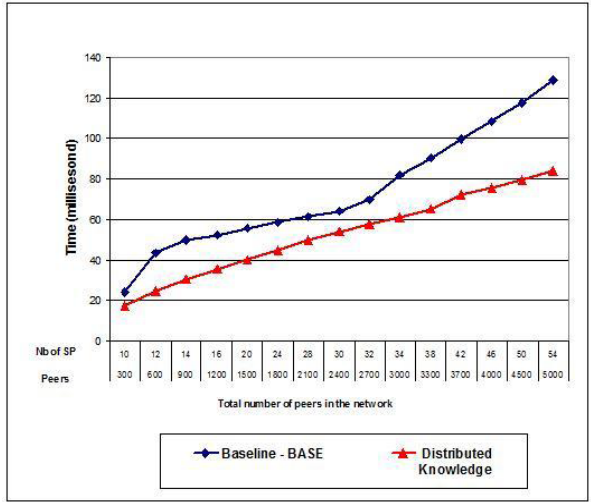

Figure 9. Response time

- The most popular measure for the effectiveness of our systems is the precision and recall.

$$precision = \frac{\# \: of \: relevant \: Responses \: retrieved}{total \: \# \: of \: retrieved \: Responses}$$

$$recall = \frac{\# \: of \: relevant \: Responses \: retrieved}{total \: \# \: of \: relevant \: Responses}$$

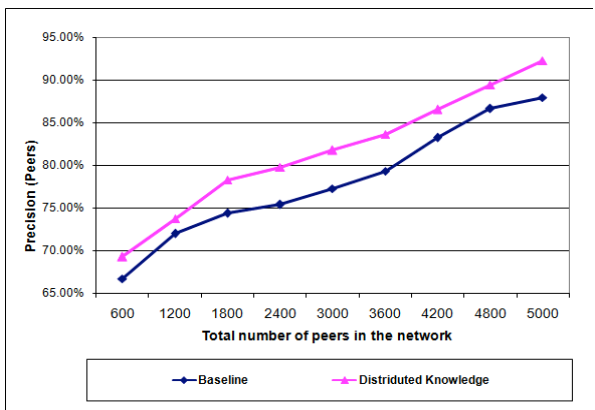

Figure 10. Precision rate

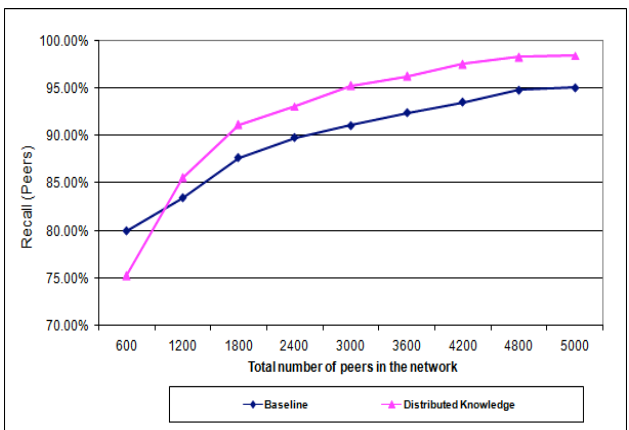

Figure 11. Recall rate

The Graphs shown in figures 9, 10 and 11 are the results of our simulations. They demonstrate the performance of using the Knowledge Super-Peer with a decision tree for routing Queries to relevant P2P domains (SP). In the first observation, the difference in the execution times between 300 and 600 peers in the DK architecture is small (See Figure 9). The execution time was measured as repository size increased.

Measurements shown in Figure 9, shows that the time increased in the DK architecture is less than the baseline architecture, in the DK architecture at 5000 Peers, the response time decreases about 35 % as for the baseline architecture, this is due to the presence of prediction mechanism in DK architecture. Measurements in Figure 9 shown the decreasing of the execution time where the number of peers and super-peers (domains) is increased. This means how much our DK architecture is scalable. Measurements in Figure 10 have shown the precision of the DK architecture compared to the Baseline architecture. In the DK architecture, we observe that the precision will increase comparing to the baseline architecture and this is due to the knowledge of all domains including in the KSP, however in the baseline architecture, we have correspondence between the neighborhood domains. This experiment was designed also to measure the accuracy of data (since precision is almost not affected completely by the network size) which is the recall (See Figure 11).The recall increases with the size of the network and reaches a percentage of almost 95 % in the DK architecture and in the baseline architecture about 91%, this is due that the baseline reduced the research space however the DK architecture increased this space research area. Finally, our Prototype in grouping P2P domains (P2P) raises some interesting performance issues. We perform experiments to demonstrate how the presence of grouping domains affects the performance and, in addition, to illustrate how grouping domains can improve the scalability of the overall system.

VII. RELATED WORK

P2P networks are quickly emerging as large-scale systems for information sharing. Through networks such as Kazaa, e-Mule, BitTorrents, consumers can readily share vast amounts of information. While initial consumer interest in P2P networks was focused on the value of the data, more recent research such as P2P web community formation argues that the consumers will greatly benefit from the knowledge locked in the data [3][4].

Query routing in a peer-to-peer network is the process by which the query is routed to a number of relevant peers and consequently it is not broadcasted on the whole network. The problem of query routing concerns the discovery of relevant peers to the query after we have denoted which peers are considered as relevant. Thus, we first have to define the criteria that make us to decide whether a peer is relevant or not. For example in some P2P systems relevant peers are these ones that match exactly all the query predicates. Secondly, we have to define the strategy on which routing will be based (e.g. based on routing indices) and all the required routing steps. Surely in peer-to-peer systems the network topology and the category of P2P determine to a large extent the applied routing strategy. Hence, before describing a routing algorithm we have to look at the characteristics of the peer-to-peer network that it will be applied to. An efficient query routing aims for limiting consuming network bandwidth by reducing messages across the network and reducing total query processing cost by minimizing the number of peers that contribute to the query's results. Finally routing in P2P networks is crucial for the scalability of the network.

Wolfgang Nejdl et. al in [5][6][7] presented the routing approach based on routing indices. This approach has been suggested and adapted under various scenarios. It is built upon an RDF-based peer-to-peer network. Queries and answers to queries are represented using RDF metadata which we can use together with the RDF metadata describing the content of peers to build explicit routing indices which facilitate more sophisticated routing approaches. Queries can then be distributed relying on these routing indices, which contain metadata information plus appropriate pointers to other (neighboring) peers indicating the direction where specific metadata (schemas) are used. These routing indices do not rely on a single schema but can contain information about arbitrary schemas used in the network. Otherwise, our approach is based on routing distributed indexes in order to find the super-peer with minimum query processing, which is the strength of our approach from approach above.

The advanced technique of [8][9] is also applied for Super-Peer Schema-Based peer-to-peer networks. Based on predefined policies a fully decentralized broadcast and matching approach distributes the peers automatically to super-peers. The basic idea here is that the super-peer establishes and maintains a specific Semantic Overlay Cluster (SOC). SOCs define peer clusters according to the metadata description of peers and their contents. Similar to the creation of views in database systems Semantic Overlay Clusters are defined by human experts. They act as virtual, abstract, independent views of selected peers in a Schema-Based P2P system. Comparing to our approach, our proposed architecture is build by regrouping the super-peers according to their interest with integrating in each group an index (decision tree) to find the relevant super-peer and other groups in an intelligent way.

Raahemi, Hayajneh and Rabinovitch [10] present a new approach using data-mining technique, in particular decision tree, to classify peer-to-peer (P2P) traffic in IP networks by capturing Internet traffic at a main gateway router, performed preprocessing on the data, selected the most significant attributes, and prepared a training-data set to which the decision-tree algorithm was applied. They built several models using a combination of various attribute sets for different ratios of P2P to non-P2P traffic in the training data. By detecting communities of peers, we achieved classification accuracy of higher than 98 percent. However, our approach uses data-mining (decision tree) to classify the super-peers (communities). By detecting communities of peers, we achieved classification accuracy of higher than 99 percent.

Bhaduri, Wolff, Giannella and Kargupta [11] propose a P2P decision tree induction algorithm in which every peer learns and maintains the correct decision tree compared to a centralized scenario. This algorithm is completely decentralized, asynchronous, and adapts smoothly to changes in the data and the network. Odysseas Papapetrou [12] proposes new approaches for enabling distributed IR over P2P without limiting the network size or mutilating the IR. The basis of these approaches is an innovative distributed clustering algorithm, which can cluster peers in a P2P network based on their content similarity. This clustering enables significant network savings and enables new families of distributed IR algorithms.

Nottelmann and Fuhr [13] build an IR system over a hierarchical P2P network. The peers there do not maintain a distributed index; instead, some super-peers are assigned the responsibility to keep their peers' summaries, and to forward the queries to the most related of their peers, or to other super-peers.

Sharma and al. [14] introduce a system, called IR-Wire, for information retrieval research in the peer-to-peer file-sharing domain. This tool maintains many statistics and implements a number of information retrieval ranking functions and contains a data logger and analyzer. The data logger logs both incoming and outgoing queries and query results and provides a way to create a snapshot of the entire data set shared by the users. The data analyzer provides a simple user interface for data analysis. This work was meant to address in the research for tools and data for P2P IR, expressed in [15].

Today's, data management in peer-to-peer (P2P) provide a promising approach that offers scalability, adaptively to high dynamics, and failure resilience. Although there exist many P2P data management systems in the literature, most of them focus on providing only information retrieval (IR) [16][17][18][19] or filtering (IF) [20] functionality (also referred to as publish/subscribe or alerting), and have no support for a combined service. DHTrie [21] is an exact IR and IF system that stresses retrieval effectiveness, while MAPS [22], [23] provides approximate IR and IF by relaxing recall guarantees to achieve better scalability.

VIII. CONCLUSION

Discovering domains on the fly are essential to perform domain directed searching. We show that while our techniques maintain the better quality of results as currently used techniques, our techniques reduce response time in P2P search (35 % at 1500 peers in DK architecture less then Baseline architecture). The advantage of our technique is the robustness in Queries routing. We experiment our technique using a Java implementation. The experiments involve communication in a large, wide-area cluster computer. We have implemented a new simulator by providing several functions many overlay protocols have in common like execution time, overlay message handling and concerning information retrieval like precision and recall. By analysis of the outcome of the experiments, we demonstrate that the system indeed shows the scalability and dependability properties predicted by our previous theoretical and simulation results. Through scalable design we have easily achieved to simulate a chord network with 5000 nodes in a reasonable amount of time. The large number of implemented overlay protocols and the availability to collect various statistical data make our simulator a powerful tool for the peer-to-peer research community. Another major direction for future work is in enhancing more the performance (Answering time) by logical restructure for our P2P network by using the minimum traverse between the super-peers (clusters). When the number of the Knowledge-super-peers increases, we jump to the logical restructure method.

ACKNOWLEDGMENT

This work has been done as a part of the project "recherche intelligente d'information multimédia multilingue arabe" by the franco-libanais comity Cedre.

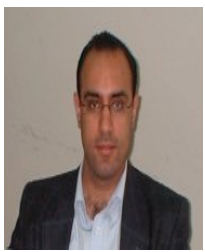
Anis ISMAIL, Born in Lebanon, April 1979, work as system and network administrator and instructor at Lebanese University, University institute of technology, saida, Lebanon. He received a PhD in Computer Science from the Aix-Marseille University. He received the B.S. degree in Telecommunication and Networking Engineering from Lebanese University (LU), the M.S. in computer science from the American University of Science and Technology (AUST) in Lebanon.

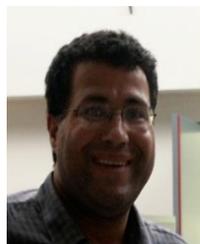
Mohamed QUAFAFOU did his PhD Thesis in 1992 on Intelligent Tutoring Systems at INSA de Lyon, France. From 1992 to 1994, he was ATER at INSA de Lyon and than at Nantes Faculty of Sciences. From 1995 to 2001, he was assistant professor at the Nantes University. During that period, he developed research on Rough Set Theory, concepts approximation, data mining, web information extraction and participated actively with France Telecom to the project Comminges to design a new web system dedicated to French web analysis for discovering emergent web communities. He was also chief-scientist at GEOBS where he headed the Geobs Data Analyzer project, which was developing a spatial data mining systems with application to environment, marketing, social analysis, etc. From September 2002, he was professor at the Avignon University and moved in 2005 to the Aix-Marseille University where he joined the Information and System Science Laboratory (UMR CNRS 6168) and continue his research on web data mining considering different application contexts like P2P, multimedia and web services. Since 2002, he teaches foundations of data/knowledge based systems including machine learning, data mining, personalization, datawarehousing, XML, web services, multimedia, web and mobile applications.

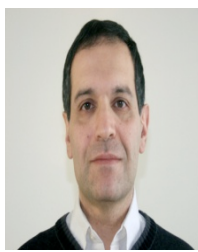
Gilles NACHOUKI is assistant professor at Nantes University in France. He received a PhD in Computer Science from the University of Toulouse. His interest domain concerns distributed databases. Currently his principal works lie in the domain of data management in peer-to- peer systems.

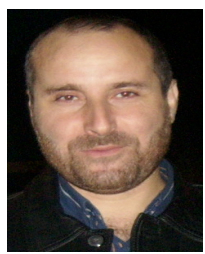
Mohammad HAJJAR is a Professor at University Institute of Technology, Lebanese University, in Lebanon. He received a PHD in computer Science at Nantes University in France. His Interest domain concerns Arabic language processing, multimedia information research and data management in peer-to- peer systems.